\documentclass[3p,times,twocolumn]{elsarticle}

\usepackage{ecrc}
\usepackage{amsmath}

\volume{00}

\firstpage{1}

\journalname{Nuclear Instruments and Methods A}

\runauth{}

\jid{nima}

\jnltitlelogo{Nuclear Instruments and Methods A}

\usepackage{amssymb}
\usepackage{hyperref}
\usepackage[figuresright]{rotating}

\begin{document}

\begin{frontmatter}



\dochead{}

\title{Simulation of the Signal Propagation for Thin-gap RPC in the ATLAS Phase-II Upgrade}

\author[a,b]{Zirui Liu}
\author[a,b]{Xiangyu Xie}
\author[a,b]{Pingxin Zhang}
\author[a,b]{Chunhao Tian}
\author[a,b]{Yongjie Sun\corref{cor1}}
\ead{sunday@ustc.edu.cn}
\cortext[cor1]{Corresponding author}
\address[a]{State Key Laboratory of Particle Detection and Electronics, University of Science and Technology of China, Hefei 230026, China}
\address[b]{Department of Modern Physics, University of Science and Technology of China, Hefei 230026, China}

\begin{abstract}
Thin-gap Resistive Plate Chambers (RPCs) with a 1 mm gap size are introduced in the ATLAS Phase-II upgrade. Smaller avalanche charge due to the reduced gap size raises concerns for signal integrity. This work focuses on the RPC signal propagation process in lossless conditions, and an analytical study is implemented for the ATLAS RPC. Detector modeling is presented, and the simulation of the RPC signal is discussed in detail.
Simulated characteristic impedance and crosstalk have been compared with the measured value to validate this model.
This method is applied to different RPC design
geometries, including the newly proposed $\eta-\eta$ readout scheme.
\end{abstract}


\end{frontmatter}

% \linenumbers

\section{Introduction}
\label{sec:intro}
RPCs are 
widely used in high-energy physics experiments as triggering detectors for muons \cite{santonico1981development}. In practice, one muon in a collision event could trigger 
 more than one readout channel due to attenuation\cite{Xie2021}, crosstalk \cite{riegler2002signal} and charge sharing \cite{lippmann2004detailed}, which harms spatial resolution and also introduce larger uncertainty to trigger algorithms \cite{Aielli:2668392}. This study aims at minimizing triggered channels in RPC readout, the signal multiplicity, through the analysis of signal propagation.

 In the ATLAS experiment, three features have been adopted to reduce the multiplicity \cite{MuonPhase2}. First 
 is the standard of the graphite layer's surface resistivity to be more than 600\ k$\Omega / \square$, which is 
 verified to make the multiplicity independent of threshold and high voltage \cite{Xie2021}.
Secondly, an extra guard strip is placed between neighboring strips. 
Thirdly, the strips must be terminated appropriately to reduce reflections.

The lossless Multi-conductor Transmission Line (MTL) theory is introduced to simulate the signal propagation process. The impact of guard strips is investigated in this paper, and the impedance of readout strips is simulated to guide termination.

In this study, an ATLAS RPC prototype, presented in Section~\ref{sec:geometry}, is modeled and measured to illustrate the analyses of signal propagation, crosstalk, and reflection. Section~\ref{sec:simu} presents the simulation, and Section~\ref{sec:meas} presents the impedance and crosstalk measurements as validation of this model. Discussion of the signal propagation of different readout panel geometries is presented in Section~\ref{sec:disc}, and the mathematical details of the lossless MTL theory for RPC are given in \ref{sec:sample:appendix}.
 
\section{Detector geometry}
\label{sec:geometry}
\begin{figure}[ht]
    \centering
    \includegraphics[width=0.43\textwidth]{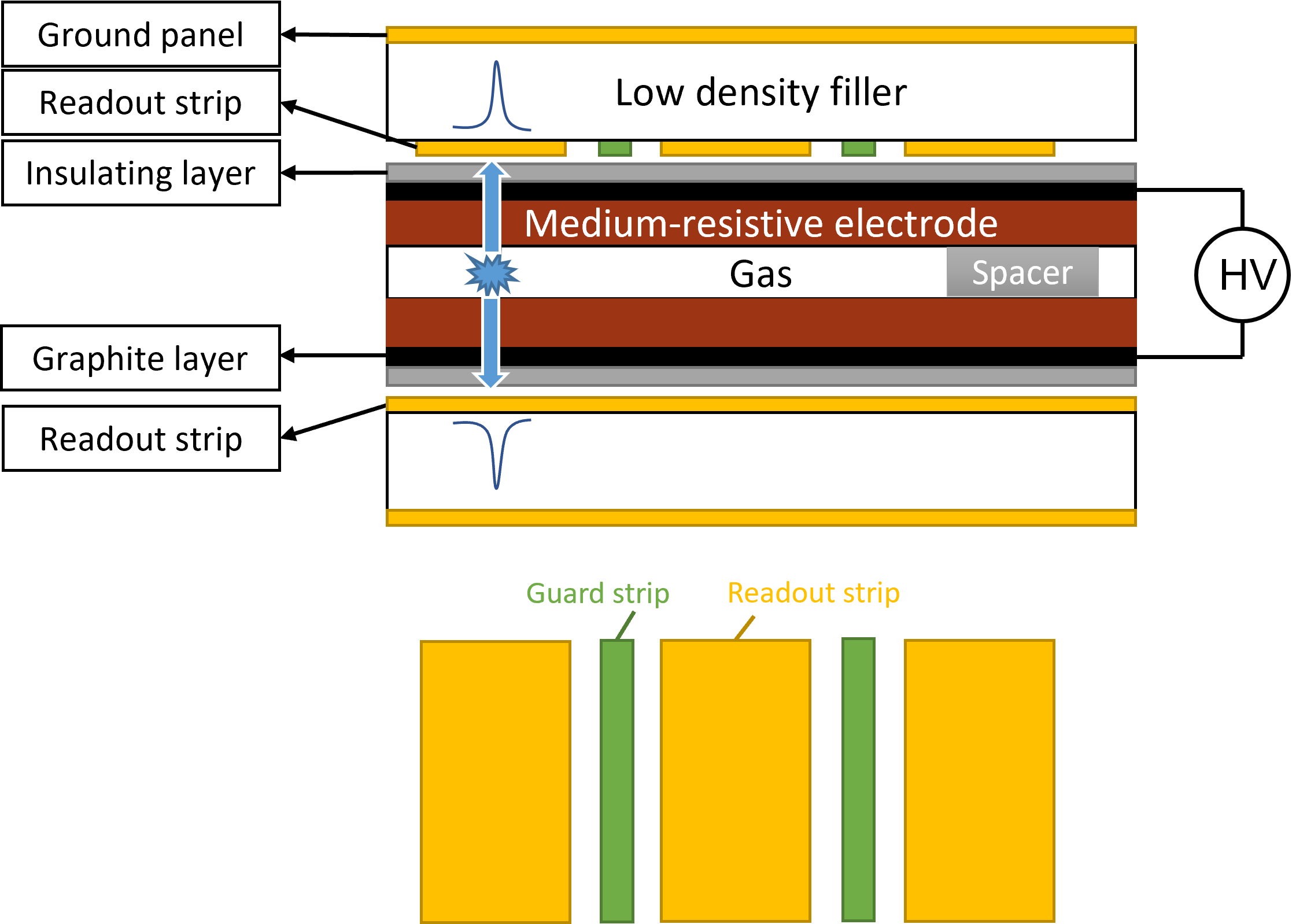}
    \caption{Detector side view and zoom in of the strip plate layout. The strip pitch is 27 mm with a 25 mm strip width and a 2 mm gap, and there is a 0.8 mm guard strip in each gap.}
    \label{fig:side}
\end{figure}

The prototype RPC used in this study is one of the four ATLAS Muon Spectrometer components (RPC, MDT, TGC, NSW) and has the same essential design parameters as the detectors to be installed at the ATLAS Phase-II upgrade. RPCs of typical designs consist of one gas gap sandwiched in-between two readout panels, as shown in Figure~\ref{fig:side}.

The gas gap consists of two pieces of medium-resistive electrodes made of bakelite separated by spacers, the graphite coating, and two PET films as the insulating layer. The operation of this detector is under 6000V nominal voltage with a gas mixture of 94.7\% of tetrafluoromethane ($\rm C_2H_2F_4$), 5.0\% of iso-butane (Iso-$\rm C_4H_{10}$) and 0.3\% of sulfur hexafluoride ($\rm SF_6$).

The readout panel is composed of a ground plate and a strip plate. Both are made of a 0.05 mm thick copper layer, and are separated by low density foam. Layout of the strip plate  is also shown in Figure~\ref{fig:side}. At the strip end, matching resistors and the front-end electronics are welded, by convention, to read out the signals. However, in this work focusing on signal propagation, the electronics are substituted by a cable to collect the signals.

The detector components parameters are measured and listed in Table~\ref{tab:layers} which are taken as input for the simulation.

\begin{table}[ht]
    \small
    \centering
    \begin{tabular}{|c|c|c|c|}
    \hline
                 Component & Material   & Thickness [mm]     &$\varepsilon_r$                              \\
            \hline
                 Gas gap & Gas mixture & 1.0 &1.0\\
                 Electrode & Bakelite & 1.2 & 5.2\\
                 Insulating layer & PET & 0.2 & 3.7\\
                 Extra air layer & Air & 0.15 & 1.0\\
                 Low density filler & Foam & 3.1 & 1.0\\
                 Ground/Strip Plates & Copper & 0.05 & N/A \\
            \hline
        \end{tabular}
    \caption{RPC detector component geometry and material characteristics.}
    \label{tab:layers}
\end{table}

\section{Simulation}
\label{sec:simu}
As mentioned in Section~\ref{sec:intro}, the surface resistivity of the graphite layer in this model is high enough to suppress the charge diffusion during propagation significantly  \cite{Xie_2019}\cite{Battistoni1982}. Along with the fact that copper has negligible resistance, the Multi-conductor Transmission Line theory is adopted in a lossless form.

The lossless MTL solution can be achieved with two basic formulae of signal propagation \cite{Riegler2000}
\begin{equation}
    \frac{d ^2}{d z^2} \vec{V}(z,t)=\hat{L} \cdot \hat{C} \frac
        {d ^2}{d t^2}\vec{V}(z,t)
        \label{equ:wave}
\end{equation}
\begin{equation}
    \frac{d ^2}{d z^2} \vec{I}(z,t)=\hat{C} \cdot \hat{L} \frac
        {d ^2}{d t^2}\vec{I}(z,t)
\end{equation}
where $\vec{V}$ and $\vec{I}$ are the voltage and current of all strips listed as vectors, and the $\hat{C}$ and $\hat{L}$ are the capacitance and inductance matrices of the transmission line system. The dimension is
  determined by the number of lines, which is the number of parallel readout 
  strips in the case of RPCs. 

In this work, the induced signal is treated as a single current source on
 the fired strip at the fired point. Another boundary condition is the 
  termination at the strip end, determined by the matching resistor network. 
For a detailed description of the math, see~\ref{sec:sample:appendix}.

The next step is to construct the $\hat{C}$ and $\hat{L}$ matrices. Considering the complex structure of the RPC readout panels and gas gap altogether, a Finite Element Method (FEM) software MAXWELL-2D tool \cite{maxwell} is used. The maximum length of each element in FEM calculation is set to be 0.2 mm, which is 1/3 of the size of the smallest structure -- the 0.6 mm gap between a strip and the adjacent guard strip.

The permittivity of the materials is re-examined to ensure the accuracy of capacitance simulation. A dielectric constant $\varepsilon_r(bakelite)=5.2$ is measured by inserting a piece of bakelite between two copper plates and then measuring this unit's capacitance. Spacers take up no more than 1\% of the area, so they are neglected in simulation, and the filler layer is regarded as a vacuum for dielectric considerations.
An extra air layer, which is not explicitly specified in detector designs, is introduced between readout strips and the insulating film. It is formed by the spacer indicators, which are 0.15 mm thick paper stickers on the surface of the gas gap, serving as an additional marker of the spacers.

The $\hat{C}$ and $\hat{L}$ matrices are solved after the convergence of field calculation in MAXWELL, and they are semi-diagonal with only leading-order and sub-leading-order elements. Higher order corrections are not considered, as the mutual capacitance of the second neighbor strips is less than ${10}^{-6}$ of the diagonal elements. 

 The matrices are three-dimensional to exploit all relevant elements, where the middle one corresponds to the fired strip,  as shown in Eq.\ref{equ:cl}.
\begin{equation}
\small
    \hat{C} = 
    \begin{pmatrix}
        229.5 & -12.4 &0\\
        -12.4 & 229.5 & -12.4 \\
        0 & -12.4 & 229.5 
    \end{pmatrix}
    \hat{L} = 
    \begin{pmatrix}
        76.2 & 2.6 & 0 \\
        2.6 & 76.2 & 2.6 \\
        0 & 2.6 & 76.2
    \end{pmatrix}
    \label{equ:cl}
\end{equation}

Unit of matrix $\hat{C}$ is pF/m and $\hat{L}$ is nH/m, and the definitions are presented in Eq.~\ref{equ:cmatrix} and Eq.~\ref{equ:lmatrix}. Following these two matrices, we can derive the propagation speed matrix.
\begin{equation}
\small
    \hat{v} = 
    \begin{pmatrix}
        236&0&0\\
        0&238&0\\
        0&0&241
    \end{pmatrix}
    [\textrm{mm/ns}]
    \label{equ:v}
\end{equation}
The average value of 238.3 mm/ns is consistent with the previously measured value of 233 mm/ns considering the uncertainty of 3\% \cite{Xie2021}. The transmission speed is a crucial parameter in novel readout methods such as double-end readout or $\eta-\eta$ readout \cite{Li2021}.
From the speed matrix, the modal dispersion can be calculated, which is at the level of $\delta v / v \simeq 0.9\%$.
Also, the impedance matrix is given in Eq.\ref{eq:zc}.
\begin{equation}
\small
\label{eq:zc}
    \hat{Z} = 
    \begin{pmatrix}
        18.3&0.8&0.04\\
        0.8&18.3&0.8\\
        0.04&0.8&18.3
    \end{pmatrix}
    [\Omega]
\end{equation}
The commonly used impedance value is the diagonal element 18.3 $ \Omega$. The matching resistor at the strip end should be close to this value to minimize signal reflection. Based on the impedance matrix, the ideal matching resistor network is given by Eq.\ref{eq:rr}.
\begin{equation}
\label{eq:rr}
\small
    \hat{R} = 
    \begin{pmatrix}
        20&400&400,000\\
        400&20&400\\
        400,000&400&20
    \end{pmatrix}
    [\Omega]
\end{equation}
From this matrix, we can read that the perfect termination requires a 20 $\Omega$ matching resistor connected to the ground, a 400 $\Omega$ resistor connecting neighboring strips, and a 400 k$\Omega$ resistor for the second neighbors. In practice, the leading-order termination (to ground) is good enough, so the complexity of higher-order termination is unnecessary. The termination network also decides the voltage detected at the front end by
    $\vec{V}(0) = \hat{R} \cdot \vec{I}(0)$.

\begin{figure}
    \centering
    \includegraphics[width=0.46\textwidth]{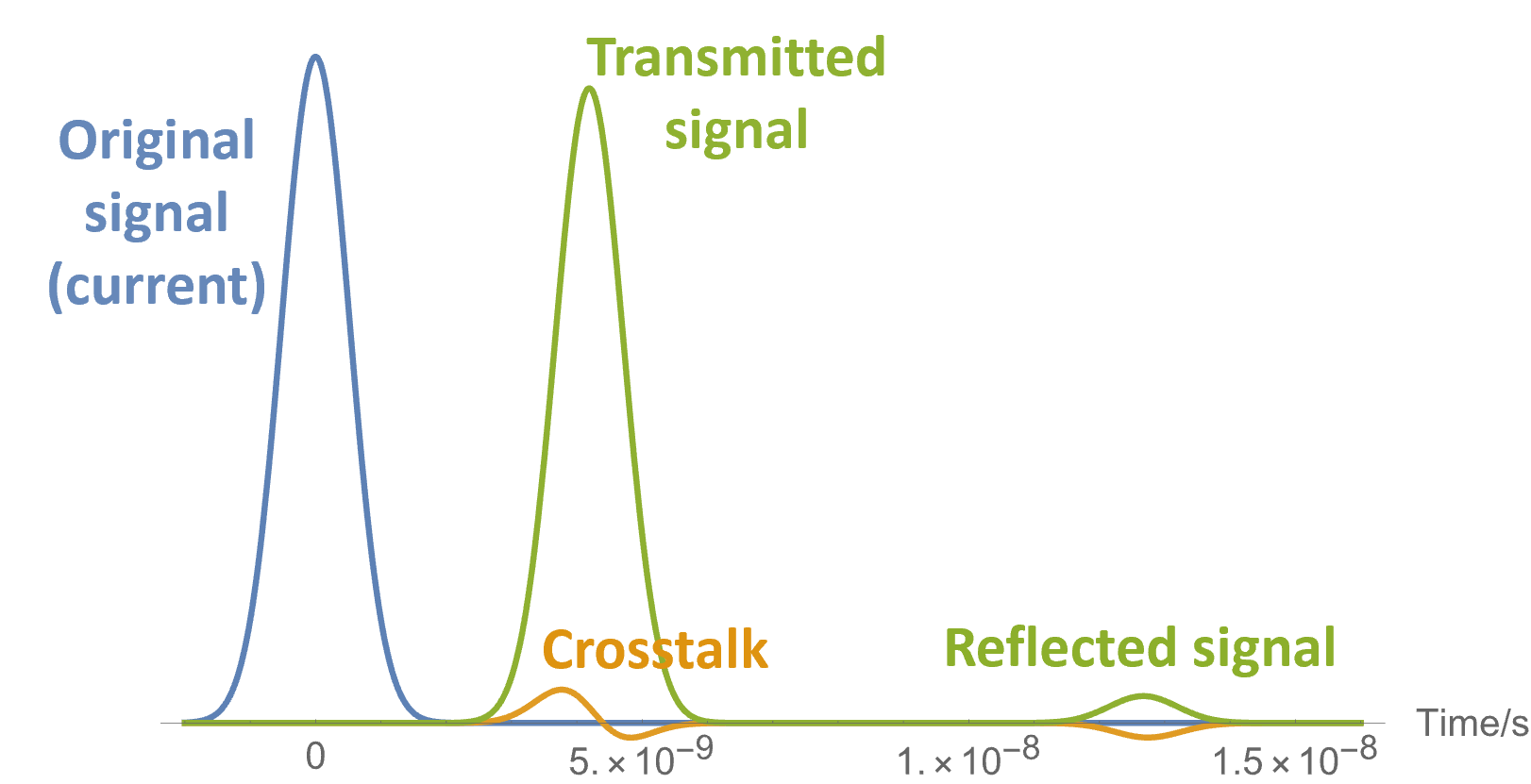}
    \caption{Time-domain solution of currents obtained with Mathematica. The propagation length is 1 m. The crosstalk signal on the neighboring strip behaves as a derivative of the transmitted signal on the fired strip. The relative amplitude of crosstalk compared to the transmitted signal is 6\%. Reflection is illustrated by using a 20 $\Omega$ matching resistor, which is higher than the impedance value of 18.3 $\Omega$. The reflected signal amplitude is 3\% of the transmitted signal.}
    \label{fig:math}
\end{figure}

The Wolfram Mathematica \cite{mathematica} software is used to give the time-domain solution with crosstalk and reflection included, as shown in Figure~\ref{fig:math}. The input signal is assumed to be a gaussian pulse with FWHM 1.22 ns, the previously measured average signal width \cite{Xie2021}. It is simulated that crosstalk emerges during propagation due to signal modal dispersion. Moreover, reflection can be calculated after assigning the value of the matching resistor.

\section{Experimental validation}
The precision of this method is dominated by the matrix elements in Eq.~\ref{equ:cl}, so the validation includes two parts to check the leading-order elements and sub-leading-order elements, respectively. The former is presented as the impedance validation in Section~\ref{sec:leading}, and the latter is presented as the crosstalk validation in Section~\ref{sec:ct}.
\label{sec:meas}
\subsection{Leading-order: impedance}
\label{sec:leading}
A Time Domain Reflectometry (TDR, Tektronix DSA8300) directly measures the impedance of the strip by the cable welded at the strip end. The impedance of the cable line is 50 $\Omega$, and the far end of the strip is grounded.

Two scenarios are validated, respectively. First, the fully assembled RPC is caged in an aluminum box, with 110 $\textrm kg/m^2$ pressure applied to ensure the uniformity of the detector and, thus, the impedance.
Second is the bare readout panel alone in the aluminum box, which provides another validation example. Both are listed in Table~\ref{tab:imp}.
\begin{table}[ht]
\small
    \centering
    \begin{tabular}{|c|c|c|}
    \hline
         Impedance [$\Omega$]&  Bare strip&Assembled\\
    \hline
         Measurement&32.2&18.5\\
         Simulation&32.3&$18.3\pm 0.4$\\
    \hline
    \end{tabular}
    \caption{Validation of impedance simulation in two scenarios. Simulation values and the systematic uncertainty for the assembled case are listed. The measured values are in good agreement with the simulation.}
    \label{tab:imp}
\end{table}

The systematic uncertainty for the assembled case is mainly contributed by the non-uniformity of the extra air layer of real detectors, for which a maximal deviation of 0.05 mm is estimated. This deviation introduces an uncertainty of 0.4 $\Omega$ to simulated impedance.

\subsection{Sub-leading-order: crosstalk}
\label{sec:ct}
The strip impedance is not sensitive to the off-diagonal elements 
when they are relatively small. However, crosstalk is directly linked with them, so the sub-leading order validation of the matrices is performed through a crosstalk study. The crosstalk is evaluated by comparing the collected signal amplitude at the fired channel and the neighboring channel.

S-parameters are introduced to quantify the level of crosstalk, defined in the frequency domain and thus independent of particular signal shape. The classical definition of S-parameter includes the phase and amplitude information so that it is expressed as a complex function \cite{159877}. For the convenience of experimental validation, only the amplitude information is preserved. In Eq.~\ref{equ:sparameter}, $A_{2}$ is the amplitude of the crosstalk, and $A_{1}$ is the transmitted signal amplitude. The S-parameter fraction is 0 when the frequency is below a characteristic frequency $f_c$ of the transmission line system.
\begin{equation}
\small
\label{equ:sparameter}
    S(f) = \frac{A_{2}(f)}{A_{1}(f)}
\end{equation}

\begin{figure}
    \centering
    \includegraphics[width=0.46\textwidth]{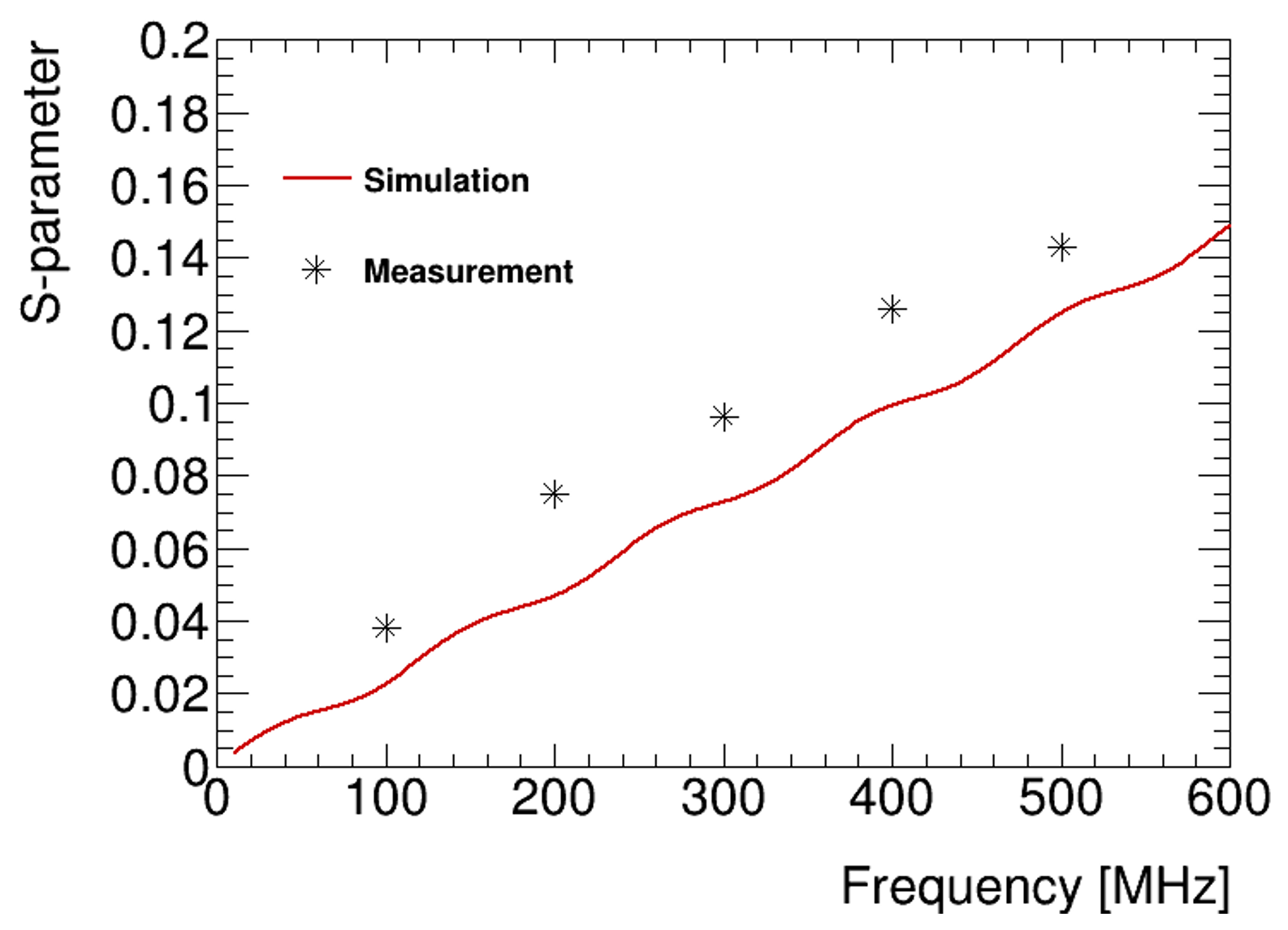}
    \caption{Comparison of measured and simulated S-parameter. The 100-500 MHz is chosen to match the typical RPC signal frequency range. The simulated S-parameter differs from the measurements by about 30\% and the trend in the dependence of frequency is described fairly well.}
    \label{fig:sp}
\end{figure}

The simulation results and the measurement of the ATLAS RPC's S-parameter fraction are shown in Figure~\ref{fig:sp}.
The simulation of S-parameters is acquired by performing Fourier transformation to the time-domain signal solutions. The measurement method is injecting sine waves at one end of the strips by a waveform generator, then collecting signals at the other end by an oscilloscope. Five points in the typical RPC signal frequency range are measured to present the difference between simulation and measurement.

\section{Discussion}
\label{sec:disc}
With the constructed modeling, different readout schemes can be tested. Implications for different geometries are also discussed as follows.

\subsection{Different geometries}
\label{sec:diff}
In various RPC geometry designs, this model applies to simulating the impedance and the crosstalk.

In barrel-shaped experiment geometry, RPCs at the outer layer usually have a larger strip pitch to maintain a consistent angular resolution. For the ATLAS RPC, the pitch of the readout strips varies from 20 to 35 mm, and this work suggests that different matching resistors should be used for different pitch for good termination. The simulation results are shown in Table~\ref{tab:pitch}.
\begin{table}[ht]
\small
    \centering
    \begin{tabular}{|c|c|c|}
    \hline
        Pitch [mm] &  Impedance [$\Omega$]&Crosstalk amplitude \\
        \hline
         35&14.4&3.8\%\\
         31&16.0&4.2\%\\
         27&18.3&6.0\%\\
         23&21.2&6.8\%\\
         20&23.4&7.5\%\\
         \hline
    \end{tabular}
    \caption{The impedance and crosstalk amplitude of the ATLAS RPC at various pitch. For wider strips, the impedance is smaller; thus, the correct matching resistor should also be smaller. The crosstalk decreases for wider pitch due to reduction of modal dispersion.}
    \label{tab:pitch}
\end{table}

Changes in the gas gap geometry also influence the impedance and the level of crosstalk. When the thickness of the bakelite electrode increases from 1.2 mm to 1.4 mm,  the impedance would be 18.7 $\Omega$ instead of 18.3 $\Omega$, and the crosstalk fraction would be 7\% instead of 6\%.

\begin{figure}
    \centering
    \includegraphics[width=0.46\textwidth]{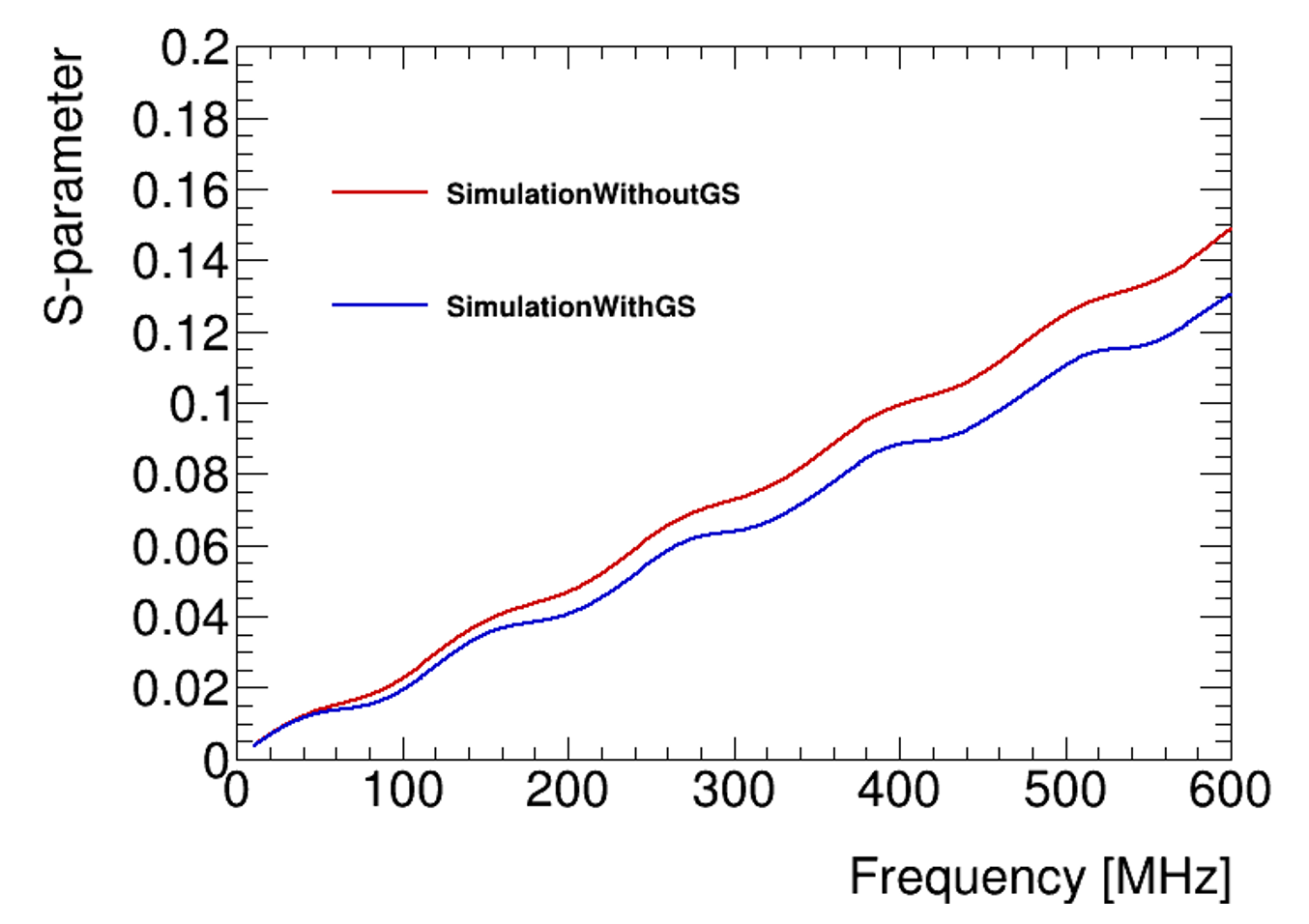}
    \caption{Comparison of the S-parameter with or without the guard strips is simulated. A drop of crosstalk fraction in the frequency-domain is observed with guard strips installed. For a typical RPC signal, crosstalk amplitude fraction would drop from 6\% to 5\% after installing guard strips. This reduction is not significant, as verified in the experiments of previous studies \cite{Ammosov2000}, though the geometry is not identical.}
    \label{fig:guard}
\end{figure}

Guard strips are implemented for the ATLAS RPC, and their impact can be investigated using the lossless MTL theory. It is quantified by computing the reduction of modal dispersion. As shown in Figure~\ref{fig:guard}, after installing guard strips, the S-parameter fraction decrease because the modal dispersion drops from 0.9\% to 0.3\% for strip lines. In addition, for guard strip lines, the simulated impedance is 104 $\Omega$, which supports the choice of 100 $\Omega$ matching resistor for guard strips in previous ATLAS RPCs \cite{MuonPhase2}.

\subsection{$\eta-\eta$ readout method}
\label{sec:eta}
\begin{figure}
    \centering
    \includegraphics[width=0.46\textwidth]{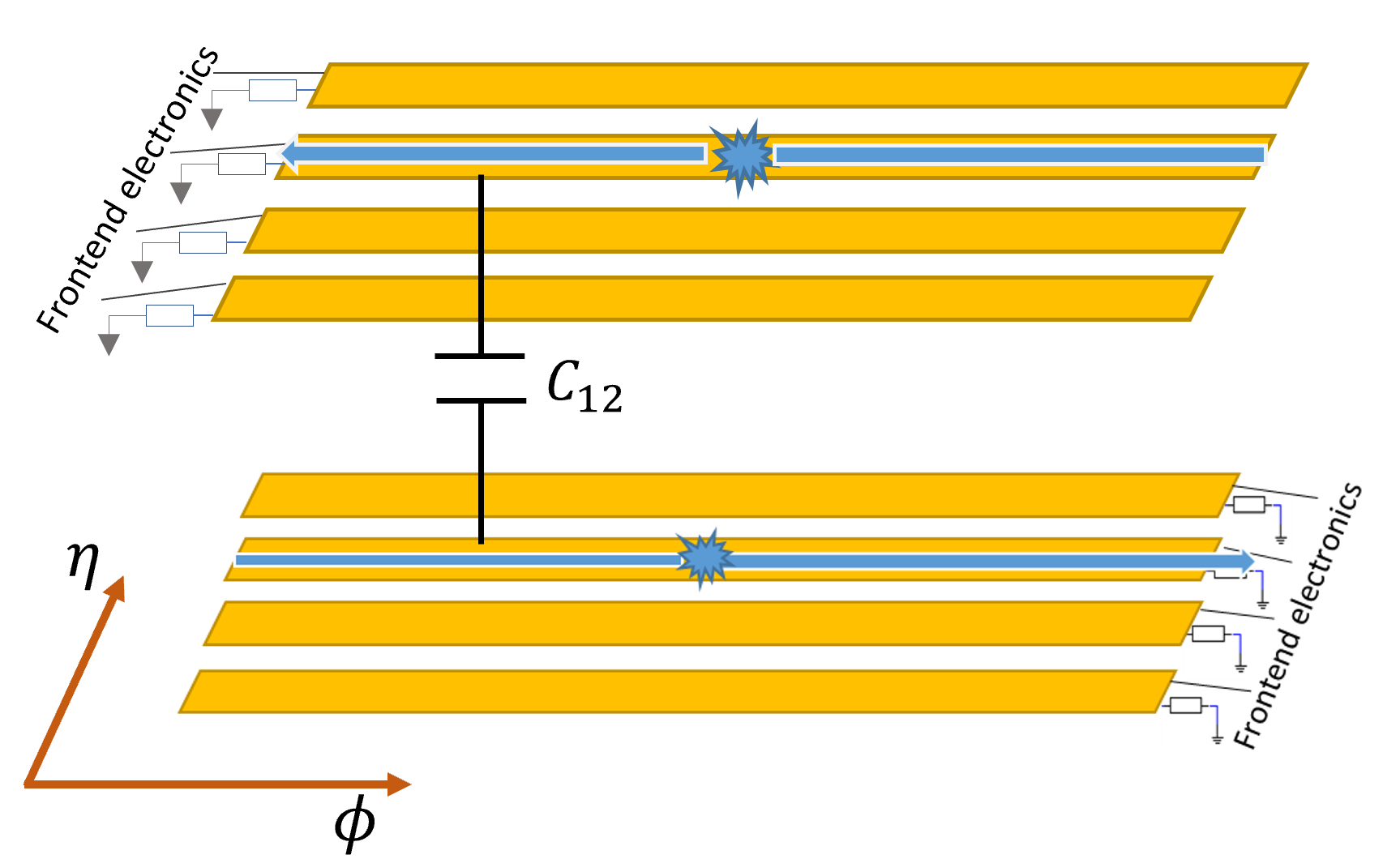}
    \caption{Illustration of $\eta-\eta$ readout method. The face-to-face strips have a mutual capacitance $C_{12}=115 \ pF/m$, nearly equal to the capacitance to ground. The large off-diagonal element in $\hat{C}$ matrix means that the "sub-leading-order" effect is no longer small compared to the leading order. As a result, sub-leading-order termination is necessary for this readout scheme.}
    \label{fig:etaeta}
\end{figure}

The traditional 2-D space-point measurement with RPC is realized by placing two orthogonal readout panels at two sides of the gas gap, which is usually referred to as $\eta-\phi$ readout in ATLAS. It is recently proposed for the ATLAS RPC Phase II upgrade that two sets of parallel strips, placed at the two sides of the gas gap, can also achieve 2-D spatial readout by comparing the time difference of the collected signals from opposite ends. This scheme is called $\eta-\eta$ readout method, as shown in Figure~\ref{fig:etaeta}. MAXWELL 2-D can simulate this scheme because the structure is homogeneous along the propagation direction.

The "face-to-face" capacitance $C_{12}$ and inductance $L_{12}$ introduce large off-diagonal elements in the impedance matrix, which is comparable with the diagonal element, as indicated in Eq.~\ref{equ:zr}. This suggests that the leading-order termination is insufficient.
\begin{equation}
\small
\label{equ:zr}
    \hat{Z}=\begin{pmatrix}
        22&10\\10&22
    \end{pmatrix}
    [\Omega]
\ \ 
    \hat{R}=\begin{pmatrix}
        32&42\\42&32
    \end{pmatrix}
    [\Omega]
\end{equation}
It is necessary to implement sub-leading-order termination, which includes not only a 32 $\Omega$ grounding resistor, but also a 42 $\Omega$ interconnect resistor between face-to-face strips. If only leading-order termination is applied, the amplitude of the reflected signal would be over 30\% of the transmitted signal, which may deteriorate the detector performance.

\section{Summary}
In this paper, a simulation method for RPC signal propagation based on lossless MTL theory is built and validated. Impedance simulation is consistent with TDR measurement, and the crosstalk is analyzed with S-parameter. RPC modeling shows good agreement with actual detectors.

Applying this model to the ATLAS RPC, the impedance of readout strips in different geometries is evaluated and guidelines in choices of matching resistors are discussed. The potential necessity of sub-leading-order termination is raised for the $\eta-\eta$ readout scheme. 

\section*{Acknowledgements}
This work is partially supported by the National Natural Science Foundation of China (No. 11961141014 and 11975228).

The authors sincerely appreciate the delightful discussions with Professor Yanwen Liu, Professor Lailin Xu and Professor Toni Baroncelli.

\appendix
\section{Lossless MTL Theory for RPC}
\label{sec:sample:appendix}
In this appendix, we present briefly the lossless MTL theory used in this work. Detailed theory can be found in \cite{paul2007analysis}.

The lossless approximation is applied so that resistance and charge sharing is not considered for this context, so that the $\hat{C}$ and $\hat{L}$ matrices are the basic elements.
\begin{equation}
\small
    \hat{C} \ =\ 
        \begin{pmatrix}
            \sum _{i=1}^N C_{1i} & -C_{12} &.. &-C_{1N}\\
            -C_{21} & \sum_{i=1}^N C_{2i}&..&-C_{2N}\\
            ..&..&..&..\\
            -C_{N1}&-C_{N2}&..&\sum_{i=1}^N C_{Ni}
        \end{pmatrix}
        \label{equ:cmatrix}
\end{equation}
\begin{equation}
\small
    \hat{L} =
        \begin{pmatrix}
            L_{11}&L_{12}&..&L_{1N}\\
            L_{21}&L_{22}&..&L_{2N}\\
            ..&..&..&..\\
            L_{N1}&L_{N2}&..&L_{NN}
        \end{pmatrix}
        \label{equ:lmatrix}
\end{equation}
where $L_{ii}$ and $C_{ii}$ are the capacitance and inductance of strip $i$ to ground and $C_{ij}$ and $L_{ij}$ are the capacitance and inductance between strip $i$ and strip $j$. The two matrices are naturally symmetric. Then the transmission speed matrix is solved in an eigenvalue equation.
\begin{equation}
\small
    \hat{M}^{-1}(\hat{C}\hat{L})\hat{M}=\hat{v}^{-2}
        \label{eigen}
\ \  
    \hat{v}^{-2} \triangleq  
        \begin{pmatrix}
            v_1^{-2}&0&..&0\\
            0&v_2^{-2}&..&0\\
            ..&..&..&..\\
            0&0&..&v_N^{-2}
        \end{pmatrix}
\end{equation}
For a system with N lines, there are typically N speed modes. The level of modal dispersion is decided by the inhomogeneity of the strip surroundings. For RPC, the inhomogeneity is mainly introduced by the difference between the $\varepsilon_r$ of bakelite and the low-density filler.

Then we can calculate the impedance matrix
\begin{equation}
\small
    \hat{Z}_c = \hat{L}\hat{M}\hat{v}\hat{M}^{-1}
\end{equation}
where $\hat{v}=Diag(v_1,...,v_N)$. And the relation between $\hat{Z}$ and $\hat{R}$ is
\begin{equation}
\small
        \hat{Y}=\hat{Z}^{-1}
    \ \ \ 
        \hat{R}=
        \begin{pmatrix}
            \frac{1}{\sum_{i=1}^N Y_{1i}}&-\frac{1}{Y_{12}}&...&-\frac{1}{Y_{1N}}\\
            -\frac{1}{Y_{21}}&\frac{1}{\sum_{i=1}^N Y_{2i}}&...&-\frac{1}{Y_{2N}}\\
            ...&...&...&...\\
            -\frac{1}{Y_{N1}}&-\frac{1}{Y_{N2}}&...&\frac{1}{\sum_{i=1}^N Y_{Ni}}
        \end{pmatrix}
    \end{equation}
The reflection coefficient is defined as
\begin{equation}
\small
            \hat{T}=\frac{\hat{I}\hat{R}-\hat{Z}}{\hat{I}\hat{R}+\hat{Z}}
            \label{my}
\end{equation}
and then the total collected signal with multiple reflection included is \cite{GonzalezDiaz2011}
\begin{equation}
\small
            \begin{split}
            \vec{V}(t)=\frac{1+\hat{T}}{2} \sum_{j=0}^{\infty}\hat{T}^{2j}
            \Biggl\{ \hat{Z} \hat{M}
            \begin{pmatrix}
                M_{1n}^{-1}I_0(t-\frac{z_0+2jD}{v_1})\\
                ...\\
                M_{Nn}^{-1}I_0(t-\frac{z_0+2jD}{v_N})
            \end{pmatrix}\\
            +\hat{T}\hat{Z} \hat{M}
            \begin{pmatrix}
                M_{1n}^{-1}I_0(t-\frac{-z_0+2(j+1)D}{v_1})\\
                ...\\
                M_{Nn}^{-1}I_0(t-\frac{-z_0+2(j+1)D}{v_N})
            \end{pmatrix}
            \Biggl\}
        \end{split}
        \label{equ:main}
        \end{equation}
where $D$ is the length of the strips and $z_0$ is the initial position of the induced signal.


%% References

%%\nocite{*}
\bibliographystyle{elsarticle-num}
\bibliography{main}

%% The provided style file elsarticle-num.bst formats 
%% references in the required Procedia style





% \end{thebibliography}

\end{document}